\documentclass{article}
\usepackage{subfig}
\usepackage{comment}
\usepackage{amsmath,amssymb}
\usepackage{graphicx}

\newtheorem{remark}{Remark}

\usepackage{natbib}
\begin{document}

\title{Isogeometric Hierarchical Model Reduction for advection-diffusion process simulation in microchannels}
\author{Simona Perotto\footnote{MOX, Dipartimento di Matematica, Politecnico di Milano, Piazza L. da Vinci 32, I-20133 Milano, Italy.}, Gloria Bellini\footnote{Politecnico di Milano, Piazza L. da Vinci 32, I-20133 Milano, Italy.}, Francesco Ballarin\footnote{Dipartimento di Matematica e Fisica, Universit\`a Cattolica del Sacro Cuore, Via della Garzetta 48, I-25133 Brescia, Italy.},\\ Karol Calò$^\S$,
Valentina Mazzi$^\S$,
Umberto Morbiducci\footnote{PoliTo\textsuperscript{BIO}Med Lab, Dipartimento di Ingegneria Meccanica e Aerospaziale, Politecnico di Torino, Corso Duca degli Abruzzi 24, I-10129 Torino, Italy.}}
\maketitle

\begin{abstract}
Microfluidics proved to be a key technology in various applications, allowing to reproduce large-scale laboratory settings
at a more sustainable small-scale. The current effort is focused on enhancing the mixing process of different passive species at the micro-scale, where a laminar flow regime damps turbulence effects. Chaotic advection is often used to improve mixing effects also at very low Reynolds numbers. In particular, we focus on passive micromixers, where chaotic advection is mainly achieved by properly selecting the geometry of microchannels. In such a context, reduced order modeling can play a role, especially in the design of new geometries.
In this chapter, we verify the reliability and the computational benefits lead by a Hierarchical Model (HiMod) reduction when modeling the transport of a passive scalar in an S-shaped microchannel. Such a geometric configuration provides an ideal setting where to apply a HiMod approximation, which exploits the presence of a leading dynamics to commute the original three-dimensional model into a system of one-dimensional coupled problems.
It can be proved that HiMod reduction guarantees a very good accuracy when compared with a high-fidelity model, despite a drastic reduction in terms of number of unknowns.
\end{abstract}

\section{Introduction}
\label{sec:1}
In the last decade, microfluidics has gained increased interest, becoming a key technology in the fields of biology and medical research, with various applications such as biosensing, diagnostics, drug discovery, regenerative medicine, tissue engineering (\cite{Nguyen2019, Sackmann2014, Bhatia2014a, Davis2008}).
The strength of microfluidics relies on the possibility to substantially reduce the sample volume by using disposable miniaturized devices which allow to replace large-scale conventional laboratory instrumentation, thus reducing hardware costs, assuring low reagent consumption and high-speed analysis, while allowing huge parallelism.

In a microfluidic system, mixing of different transported passive species is a crucial issue for the optimization of chemical and biochemical reactions involved in the process of medical diagnostics, drug discovery, chemistry production and proteomics (\cite{Suh2010}). However, at the microscale, mixing is often difficult to be achieved in common practice, since microfluidics is characterized mainly by very low Reynolds numbers (ranging from less than unity up to a few hundreds), and cannot take advantage of turbulence to improve mixing efficiency. Under laminar flow, mixing is mainly a diffusion-driven phenomenon characterized by long time scales and high diffusive lengths, which can become prohibitive in view of a numerical modeling.

To overcome these limitations, great efforts have been made to develop new techniques in order to achieve rapid and efficient laminar flow mixing in microsystems (\cite{Hardt2005a}). In particular, chaotic advection can be exploited to enhance the diffusive process in microfluidics. Chaotic advection is obtained by stretching and folding the interface between miscible streams, thus reducing the diffusive length, and improving the mixing efficiency, also at very low Reynolds numbers. In passive micromixers, chaotic advection is usually achieved by properly designing the three-dimensional geometry of microchannels (\cite{846699,Lin2007}).
It is well known that in a curved channel the flow undergoes a centrifugal displacement of the maximal axial velocity (\cite{doi:10.1080/14786440408564513}), leading to the onset of secondary flows, which in turn enhance mixing.
Thus, several micromixer designs are based on curved microchannels, such as spiral microchannels (\cite{Sudarsan2006}), U-shaped microchannels (\cite{Gigras2008}), or clothoid-based geometries (\cite{Pennella2012}).

Following this principle, in this study an S-shaped microchannel with fixed curvature is considered, in the simplest case of a single passive species. In such a way, we recover the conventional setting to apply a Hierarchical Model (HiMod) reduction. As a matter of fact, HiMod proved to be an effective tool (\cite{GuzzettiPerottoVeneziani18,brandesperotto2020}) to model phenomena exhibiting a main dynamics (here represented by the advection of the passive scalar along the microchannel), in the presence of local secondary dynamics evolving along the transverse sections (in the case of interest, represented by the mixing effects induced by the geometry).
Analogously to other model reduction procedures (we refer, e.g., to~\cite{ChinestaKeuningsLeygue13,GonzalesetAl10} and to~\cite{PerottoCarlinoBallarin2020} for a comparison between HiMod and another well-established model reduction technique), a HiMod reduction exploits a standard separation of variables and describes the mainstream and the secondary dynamics with different approximation schemes. In the first proposal, the mainstream is discretized by one-dimensional finite elements, while the transverse dynamics are modeled by a modal expansion (\cite{ern2008hierarchical,perotto2010hierarchical,PerottoZilio13,perotto2014survey}). For the specific context at hand, we resort to the variant of the original approach which employs an isogeometric discretization along the leading direction, in order to deal with generic geometries (\cite{PerottoRealiRusconiVeneziani17,brandesperotto2020}).

Independently of the selected discretizations, HiMod reduction considerably contains the computational effort, in particular when compared with full three-dimensional models, without quitting accuracy.
Indeed, a HiMod expansion leads to commute the full model into a system of coupled one-dimensional problems solved along the leading direction, whose
coefficients include the effect of the transverse dynamics. This ensures several computational simplifications, especially in the presence of complex geometries, primarily the computational domain we have to discretize which is  a one-dimensional instead of a three-dimensional one.
In terms of accuracy, the HiMod approximation can be arbitrarily enriched by properly increasing the number of modal functions used to describe the secondary transverse dynamics (\cite{perotto2010hierarchical,perotto2014coupled,perotto2015space,aletti2018himod}).

The paper is organized as follows. Section~\ref{sec:2} introduces the partial differential equation model used to describe the transport of the tracked species inside the S-shaped microchannel.
Section~\ref{sec:3} is devoted to set up the isogeometric HiMod reduction, by addressing (i) the geometric characterization of the computational domain, which distinguishes between the mainstream and the secondary directions; (ii) the definition of thefp discrete HiMod space; (iii) the HiMod algebraic formulation. Numerical results are discussed in Section~\ref{sec:res}, both in terms of accuracy and computational saving. Finally, Section~\ref{sec:concl} summarizes conclusions and possible future perspectives.

\section{Advection-diffusion problems in a microchannel}
\label{sec:2}
\begin{figure}[t]
    \centering
    \includegraphics[width=7.3cm]{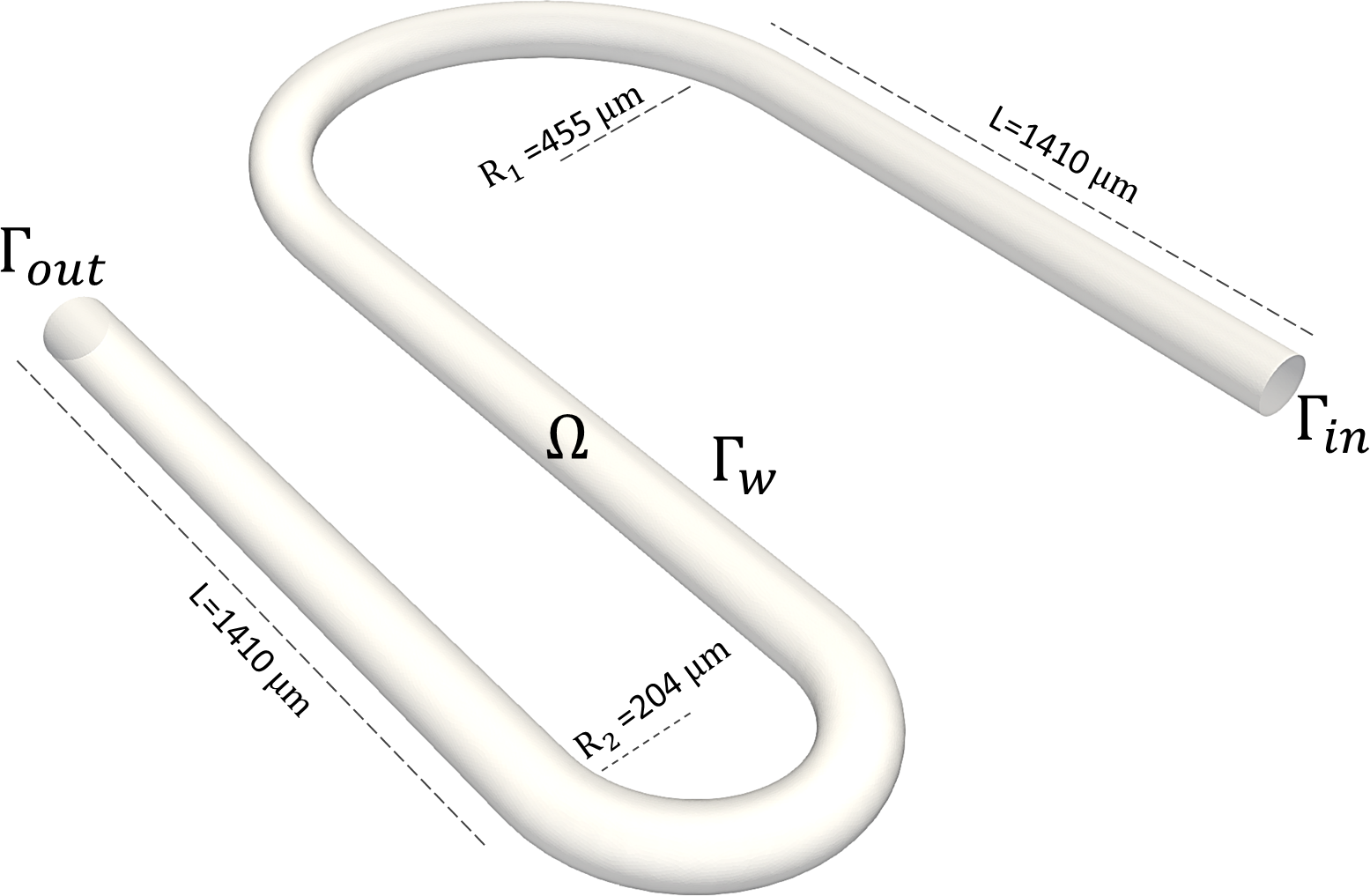}
    \caption{Geometry of the S-shaped microchannel with two fixed curvatures, $\Omega$: $\Gamma_{in}$: inlet section; $\Gamma_{w}$: lateral boundary; $\Gamma_{out}$: outlet section.}
    \label{fig1:Methods}
\end{figure}

The problem we consider models the transport of a passive scalar in the
S-shaped microchannel with two fixed curvatures, $\Omega \subset \mathbb{R}^3$,
sketched in Figure~\ref{fig1:Methods}.
To this aim, we resort to a standard advection-diffusion (AD) equation, that is formulated as\footnote{Throughout the paper, standard notation are adopted for the function spaces (\cite{ern04}).}
\begin{equation}\label{problem_adr3d}
\left\{
\begin{array}{lll}
    -\nabla \cdot (\nu \nabla u(\mathbf{z})) + \nabla \cdot (\mathbf{b}(\mathbf{z})u(\mathbf{z}))  = f(\mathbf{z}) \qquad &\text{in } \Omega \\[2mm]
    u(\mathbf{z}) = u_{in}\qquad  &\text{on } \Gamma_{in} \\[2mm]
    u(\mathbf{z}) = 0  &\text{on } \Gamma_{w} \\[1mm]
    \nu \displaystyle \frac{\partial u}{\partial \mathbf{n}}(\mathbf{z}) = 0 &\text{on } \Gamma_{out},
\end{array}
\right.
\end{equation}
where $u$ denotes the concentration of the passive scalar, the convective field $\mathbf{b}\in [L^\infty(\Omega)]^3$, with $\nabla \cdot \mathbf{b}\in L^2(\Omega)$, corresponds to the velocity of the fluid in the microchannel,
$f\in L^2(\Omega)$ models the action of an external force, $\nu \in \mathbb{R}^+$ is the diffusion coefficient of the transported species, $u_{in} \in H^{1/2}(\Gamma_{in})$ is the  concentration at the inlet $\Gamma_{in}$, with
$\mathbf{z}$ the vector collecting the spatial coordinates in $\Omega$, $\mathbf{n}$ the outer unit normal to the boundary $\partial \Omega$, $\Gamma_{in}$, $\Gamma_{out}$ the inlet and the outlet sections, $\Gamma_w$ the lateral boundary, so that $\partial \Omega=\Gamma_{in}\cup \Gamma_{w} \cup \Gamma_{out}$.\\
We observe that, since the dynamics in a microchannel is usually characterized by a very small Reynolds number, we can assume that the fluid is incompressible, namely $\nabla \cdot \mathbf{b}(\mathbf{z}) = 0$ in $\Omega$, and that the fluid flow is laminar, so that, in practice, we can compute $\mathbf{b}$ by solving a steady Navier-Stokes problem (see Section \ref{sec:res}).

In view of the application of the Hierarchical Model (HiMod) reduction technique, we introduce the weak formulation of problem \eqref{problem_adr3d}. To this end, we introduced the Sobolev spaces $H^1(\Omega) = \{v \in L^2(\Omega): \nabla v \in\ [L^2(\Omega)]^3\}$ and $H^1_{0,\Gamma_D}(\Omega) = \{v \in H^1(\Omega): v|_{\Gamma_D}=0\}$,
with $\Gamma_D=\Gamma_{in} \cup \Gamma_{w}$.
Thus, the weak formulation of the reference AD problem reads: Find $u \in V$, with $V=H^1_{0,\Gamma_D}(\Omega)$, such that
\begin{equation} \label{elliptic prob}
a(u,v)=\textit{F}(v)\qquad \forall v \in V,
\end{equation}
where $a: V \times V \longrightarrow \mathbb{R}$ and $F : V \longrightarrow \mathbb{R}$ are the bilinear and the linear forms associated with model \eqref{problem_adr3d}, namely
\begin{equation}
    a(u,v) = \int_{\Omega} \nu\ \nabla u(\mathbf{z}) \cdot \nabla v(\mathbf{z}) d\mathbf{z}\, + \int_{\Omega} \mathbf{b}(\mathbf{z}) \cdot \nabla u(\mathbf{z}) v(\mathbf{z}) d\mathbf{z}
    \label{a_ADR}
\end{equation}
\begin{equation}
    F(v) = \int_{\Omega} f(\mathbf{z}) v(\mathbf{z}) d\mathbf{z}\, - \, a(R_{u_{in}}, v),
    \label{F_ADR}
\end{equation}
with $R_{u_{in}}$ a suitable lifting of the boundary data $u_{in}$ on $\Gamma_{in}$.\\
Thanks to the assumptions introduced on the problem data, we can guarantee the well-posedness of the weak problem \eqref{elliptic prob} by applying the Lax-Milgram lemma (\cite{ern04}).

Hereafter, we will refer to \eqref{elliptic prob} as to the \textit{full problem}, the HiMod reduction will be applied to.

\section{Numerical modelling using HiMod Reduction}\label{sec:3}
In this section, we apply the HiMod reduction to the full problem \eqref{elliptic prob}, by discussing the three main steps typical of a HiMod formulation: (i) the geometric characterization of the computational domain that distinguishes between a leading and a transverse direction (Section \ref{sec:geo}); (ii) the definition of the function spaces associated with either directions and their combination in the setting of the HiMod discrete space (Section \ref{sec:spaces}); (iii) the resulting algebraic formulation, which reduces to a system of one-dimensional (1D) coupled problems (Section \ref{sec:algebraic}). For further details, we refer the reader, e.g., to \cite{ern2008hierarchical,perotto2010hierarchical,perotto2014survey}.

\subsection{Geometric characterization of the computational domain}\label{sec:geo}
The main idea behind a HiMod reduction is to distinguish in the physical domain $\Omega$ a \textit{leading direction}, associated with a 1D supporting fiber $\Omega_{1D}=[x_{in}, x_{out}]$, and a set of secondary orthogonal \textit{transverse directions}, parallel to the 2-dimensional (2D) transverse fibers $\gamma_x\subset \mathbb{R}^{2}$, at the generic point $x$ along $\Omega_{1D}$.
In other words, we are assuming to perform the reduction in a three-dimensional (3D) \textit{fiber bundle}, so that $\Omega=\displaystyle \cup_{x \in \Omega_{1D}} \{x\} \times \gamma_x$. With reference to the specific application to the microchannel in Figure~\ref{fig1:Methods}, the leading direction $\Omega_{1D}$ coincides with the centerline of the microchannel $\Omega$, $x$ is the curvilinear abscissa along $\Omega_{1D}$, and fibers $\gamma_x$ are the circular trasverse sections of the channel, orthogonal to the centerline.

In view of Section~\ref{sec:spaces}, we introduce a smooth map that changes the physical domain, $\Omega$, into a reference domain, $\hat{\Omega}$, where computations are easier and are performed once and for all. Thus, we define the invertible maps
${\Psi}: \Omega \rightarrow \hat{\Omega}$ and ${\Phi}: \hat{\Omega} \rightarrow \Omega$,  such that $\Phi(\cdot)={\Psi}^{-1}(\cdot)$.
Without loss of generality, we identify the reference domain with a unit cube.
The same directional decomposition as for the physical domain is applied to the reference domain, so that $\hat{\Omega} = \cup_{\hat{x} \in \hat{\Omega}_{1D}} \{\hat{x}\} \times \hat{\gamma}$, where $\hat{\Omega}_{1D}=[\hat x_{in}, \hat x_{out}]\subset \mathbb R$ is a unit length segment, while $\hat{\gamma}$ coincides with the unit square $(0, 1)^2$.
In particular, in Section \ref{sec:res}, following an IsoGeometric Analysis (IGA) approach (\cite{IGA_hughes}), we will employ Non-Uniform Rational B-Splines (NURBS) functions to define the map ${\Psi}$ characterizing the microchannel geometry in Figure~\ref{fig1:Methods}.

Finally, suppose that $\textbf{z} = (x,\textbf{y})$ and $\hat{{\textbf{z}}} = (\hat{x},\hat{\textbf{y}})$ denote the generic point in $\Omega$ and $\hat{\Omega}$, respectively such that ${\Psi}({\textbf{z}}) = ({\psi}_1({\textbf{z}}), {\psi}_2({\textbf{z}}))=\hat{{\textbf{z}}}$ (i.e., $\hat{x}={\psi}_1({\textbf{z}})$, $\hat{\textbf{y}}={\psi}_2({\textbf{z}})$) and ${\Phi}({\hat{\textbf{z}}}) = ({\varphi}_1({\hat{\textbf{z}}}), {\varphi}_2({\hat{\textbf{z}}}))={{\textbf{z}}}$ (i.e., ${x}={\varphi}_1(\hat{{\textbf{z}}})$, ${\textbf{y}}={\varphi}_2(\hat{{\textbf{z}}})$).
We assume maps $\Psi$ and $\Phi$ to be differentiable with respect to ${{\textbf{z}}}$ and $\hat{{\textbf{z}}}$, respectively and
we define the Jacobian of such transformations, namely,
$$
    \mathcal{J}_\Psi(\textbf{z})=\displaystyle \frac{\partial \Psi}{\partial\textbf{z}}(\textbf{z})=\begin{bmatrix}
\partial_x \psi_1(\textbf{z})& \nabla_{\textbf{y}} \psi_1(\textbf{z})\\[2mm]
\partial_x \psi_2(\textbf{z}) & \nabla_{\textbf{y}} \psi_2(\textbf{z})
\end{bmatrix}, \   \mathcal{J}_\Phi(\hat{\textbf{z}})=\frac{\partial \Phi}{\partial\hat{\textbf{z}}}(\hat{\textbf{z}})=\begin{bmatrix}
\partial_{\hat{x}} \varphi_1(\hat{\textbf{z}}) & \nabla_{\hat{\textbf{y}}} \varphi_1(\hat{\textbf{z}})\\[2mm]
\partial_{\hat{x}} \varphi_2(\hat{\textbf{z}}) & \nabla_{\hat{\textbf{y}}} \varphi_2(\hat{\textbf{z}})
\end{bmatrix}
$$
in $\mathbb{R}^{3 \times 3}$, where
$\nabla_{\mathbf{y}}$ and $\nabla_{\hat{\mathbf{y}}}$ stand for the gradient with respect to $\mathbf{y}$ and $\hat{\mathbf{y}}$, respectively.
Due to the smoothness assumptions on $\Phi$ and $\Psi$, it holds $\mathcal{J}_{\Psi}({\textbf{z}}) = \mathcal{J}_{\Phi}^{-1}({\hat{\textbf{z}}})$.

In view of Section~\ref{sec:algebraic}, all the integrals in $\Omega$ in \eqref{a_ADR}-\eqref{F_ADR} will be traced back to integrals on the reference domain $\hat{\Omega}$ by applying the change of variable formula
\begin{equation}\label{eq:changevar}
\int_{\Omega} g(\textbf{z}) d\textbf{z} =
    \int_{\Omega} g(x, \textbf{y}) d\textbf{z} = \int_{\hat{\Omega}} g({\varphi}_1(\hat{\textbf{z}}), {\varphi}_2(\hat{\textbf{z}})) \, | \text{det}(\mathcal{J}_{{\Phi}})| \, d \hat{\textbf{z}},
\end{equation}
with $g\in L^2(\Omega)$.

\subsection{The discrete HiMod space}
\label{sec:spaces}
The fiber structure introduced on $\Omega$ is instrumental in the definition of the following function spaces:
\begin{itemize}
    \item the 1D space $\hat{V}_{1D}$, spanned by functions defined on the reference supporting fiber $\hat{\Omega}_{1D}$, so that $\hat{V}_{1D} \subseteq H^1(\hat{\Omega}_{1D})$. For consistency reasons, functions in $\hat{V}_{1D}$ must be compatible with the boundary conditions enforced on $\Gamma_{in}$ and  $\Gamma_{out}$; for instance, if a homogeneous Dirichlet boundary condition is assigned on $\Gamma_{in}$, functions in $\hat{V}_{1D}$ must vanish at $\hat x_{in}$.
    In particular, to provide the discrete HiMod formulation, we identify space $\hat{V}_{1D}$ with
    a finite dimensional discrete space. In early works (see, e.g.,~\cite{ern2008hierarchical,perotto2010hierarchical}), the supporting fiber is assumed to be a segment, and a finite element (FE) basis is employed to define $\hat{V}_{1D}$. Since in the microchanel application the centerline is curved, we rely on an IGA discretization, following~\cite{brandesperotto2020};
    \item the 2D space $V_{m, \hat{\gamma}}$, spanned by functions defined on the reference fiber $\hat{\gamma}$.  In particular, we define a \textit{modal basis} of functions $\{\phi_k\}_{k \in \mathbb{N}^+} \subset H^1(\hat{\gamma})$, orthonormal with respect to the $L^2(\hat{\gamma})$-scalar product, i.e., such that
\begin{equation}
    \int_{\hat{\gamma}} \phi_k(\hat{\textbf{y}}) \phi_l(\hat{\textbf{y}}) d\hat{\textbf{y}} = \delta_{kl} \qquad \forall k,l \in \mathbb{N}^+,
    \label{orthogonality condition}
\end{equation}
with $\delta_{kl}$ the Kronecker symbol. With this modal basis, we associate the space $V_{\infty, \hat{\gamma}} = \text{span} (\{\phi_k\})$, as well as the truncated function space $V_{m, \hat{\gamma}} = \text{span} (\{\phi_k\}_{k=1}^m)$.
Consistently with the definition of space $\hat V_{1D}$, we enforce the boundary conditions assigned on $\Gamma_w$ to the modal functions. Since, in the microchannel configuration, we set homogeneous Dirichlet boundary conditions on $\Gamma_w$, the modal basis coinsists of sinusoidal functions, vanishing on $\partial \hat{\gamma}$. In Remark~\ref{rem_base}, we provide some comments about a practical way to select the modal functions in order to fulfill generic boundary data. \\
Moreover, we notice that relation \eqref{orthogonality condition}, and thus space $V_{m, {\hat{\gamma}}}$, does not actually depend on $\hat{x}$, being $\hat{\gamma}$ the unit square, independently of $\hat x$. This means that we are allowed to pre-compute all the integrals involving the modal functions, independently of the specific case study at hand.
\end{itemize}

\begin{remark}\label{rem_base}
In~\cite{aletti2018himod} the authors propose a practical method to build
a modal basis of functions which include, in an essential way, any kind (Dirichlet, Neumann, Robin, mixed) of boundary data on $\Gamma_w$. The idea is to solve an auxiliary Sturm-Liouville Eigenvalue (SLE) problem, characterized by the differential operator ${\mathcal L}_{SLE}$, on the reference fiber $\hat \gamma$, by imposing on $\partial \hat \gamma$ the same boundary condition as on $\Gamma_w$. Concerning the choice of ${\mathcal L}_{SLE}$, it is expected to be a symmetric operator. For instance, with reference to the AD problem in \eqref{problem_adr3d}, it turns out that ${\mathcal L}_{SLE}$ reduces to the diffusive operator $-\nabla \cdot (\nu \nabla)$, due to the incompressibility assumption on the fluid.
The eigenfunctions of the SLE problem
constitute the modal basis $\{\phi_k\}$, which is named \emph{educated}, since it automatically takes into account features of the solution to be reduced. For more details about computational aspects of this approach as well as for a modeling convergence analysis,  we refer the reader to the original paper.
\end{remark}

Now, by exploiting a separation of variable principle, we define the
\textit{hierarchically reduced function space}, $V_m$, obtained by combining spaces $\hat V_{1D}$ and $V_{m, \hat{\gamma}}$, as
\begin{equation}\label{global function space}
    V_m=\Big\{ v_m(\textbf{z}) = \sum_{k=1}^m \tilde{v}_k(\psi_1(\textbf{z})) \phi_k(\psi_2(\textbf{z})) \, :\, \tilde{v}_k \in \hat{V}_{1D}, \, \textbf{z} \in \Omega \Big\},
\end{equation}
where $m \in \mathbb{N}^+$ denotes the modal index, here set a priori (we refer to~\cite{perotto2014coupled,perotto2015space} for an automatic selection strategy of the modal index). Notice that, thanks to the orthonormality condition \eqref{orthogonality condition}, coefficient $\tilde{v}_k$ in \eqref{global function space} can be interpreted as the frequency associated with the $k$-th modal basis function, being
\begin{equation}
    \tilde{v}_k(\psi_1(\textbf{z})) = \int_{\hat{\gamma}} u_m(\psi_1(\textbf{z}), \psi_2(\textbf{z})) \phi_k(\psi_2(\textbf{z})) d\hat{\textbf{y}}.
    \label{frequency coefficients}
\end{equation}
Once space $V_m$ is defined, the \textit{hierarchically reduced problem} reads as follows: Fixed $m \in \mathbb{N}^+$, find $u_m \in V_m$ such that
\begin{equation}\label{reduced problem}
    a(u_m, v_m)= \textit{F}(v_m) \qquad \forall v_m \in V_m,
\end{equation}
or likewise, after picking
in \eqref{reduced problem} the test function as $v_m(\textbf{z})=\theta_l(\psi_1(\textbf{z})) \phi_k(\psi_2(\textbf{z}))$ and by exploiting
the modal representation $u_m($\textbf{z}$) = \displaystyle \sum_{j=1}^m \tilde{u}_j(\psi_1(\textbf{z})) \phi_j(\psi_2(\textbf{z}))$ for the trial function, find $\tilde{u}_j\in \hat V_{1D}$, for any $j\in \{1, \ldots, m\}$, such that
\begin{equation}
    \sum_{j=1}^m a\big(\tilde{u}_j(\psi_1(\textbf{z}))\phi_j(\psi_2(\textbf{z})),\theta_l(\psi_1(\textbf{z})) \phi_k(\psi_2(\textbf{z}))\big) = \textit{F}\big(\theta_l(\psi_1(\textbf{z})) \phi_k(\psi_2(\textbf{z}))\big),
    \label{new_problem}
\end{equation}
for any $k \in \{1,...,m\}$ and
with $\{\theta_l\}_{l=1}^{N_h}$ the set of the basis functions for space $\hat{V}_{1D}$, with $N_h={\rm dim}(\hat{V}_{1D})$ the associated dimension.\\
The practical effect of a HiMod approximation is to commute the full 3D
model \eqref{elliptic prob} into the system \eqref{new_problem} of $m$ coupled 1D problems, defined along the centerline $\hat{\Omega}_{1D}$ of the reference domain $\hat{\Omega}$.

Finally, following~\cite{perotto2010hierarchical}, we endow space $V_m$ both with a conformity ($V_m\subset V$, for any $m\in \mathbb N^+$) and with a spectral approximability ($\lim_{m\to +\infty} \inf_{v_m\in V_m} \| v- v_m\|_V=0$, for any $v\in V$) hypotheses, in order to ensure the well-posedness of the HiMod formulation \eqref{reduced problem} as well as the convergence of the HiMod approximation $u_m$ to the weak solution $u$ in \eqref{elliptic prob}, for $m\to +\infty$. This means that the accuracy of the reduced model can be arbitrarily set by suitably tuning the modal index $m$ in the reduced formulation \eqref{reduced problem}.

\subsection{The HiMod algebraic formulation} \label{sec:algebraic}
In this section we derive the algebraic counterpart of the HiMod formulation in \eqref{reduced problem}, taking, for simplicity, $u_{in}=0$ in \eqref{problem_adr3d} so that the lifting term in \eqref{F_ADR} is null.\\
With a view to the numerical verification in Section~\ref{sec:res}, we identify the basis $\{\theta_l\}_{l=1}^{N_h}$ for $\hat V_{1D}$ used in \eqref{new_problem} with the set $\{\mathcal{R}_l\}_{l=1}^{N_h}$ of the NURBS functions defined in the interval $[0, 1]$ (\cite{IGA_hughes}).
Thus, the HiMod reduced solution and the test function can be expressed as
\begin{equation}\label{expanded}
    u_m(\mathbf{z}) = \sum_{j=1}^{m} \sum_{i=1}^{N_h} u_{ij} \mathcal{R}_i(\psi_1(\mathbf{z}))\phi_j(\psi_2(\mathbf{z})), \quad
    v_m(\mathbf{z}) = \mathcal{R}_l(\psi_1(\mathbf{z}))\phi_k(\psi_2(\mathbf{z}))
\end{equation}
for $k\in \{1,\ldots,m\}$, $l\in \{1,\ldots,N_h\}$, so that
the actual unknowns of the HiMod reduced formulation are the $mN_h$ coefficients ${u}_{ij} \in \mathbb R$, with $j\in \{1,\ldots,m\}$, $i\in \{1,\ldots,N_h\}$.\\
The HiMod algebraic system can be derived by exploiting expansions \eqref{expanded} in the definition of the bilinear and linear forms \eqref{a_ADR} and \eqref{F_ADR}, combined with the pull-back transformation from $\Omega$ to $\hat{\Omega}$.
To this end, we move from
the gradient expansion
\begin{equation}
\begin{array}{lll}
    \nabla(w(\psi_1(\mathbf{z}))\phi_s(\psi_2(\mathbf{z}))) &=&
    \phi_s(\psi_2(\mathbf{z})) w'(\psi_1(\mathbf{z})) \left[ \begin{array}{c}
           \partial_{{x}}\psi_1(\mathbf{z})\\[2mm]
          \nabla_{{\mathbf{y}}}\psi_1(\mathbf{z})
    \end{array}\right]
     \\[5mm]
    &+& w(\psi_1(\mathbf{z}))
    \phi'_s(\psi_2(\mathbf{z})) \left[ \begin{array}{c}
    \partial_{{x}}\psi_2(\mathbf{z})\\[2mm]
    \nabla_{{\mathbf{y}}}{\psi_2}(\mathbf{z})\end{array}\right],
    \label{expansion_ad3_newT}
    \end{array}
\end{equation}
with $w'(\psi_1(\mathbf{z}))=dw(\psi_1(\mathbf{z}))/d \psi_1(\mathbf{z})$, $\phi'_s(\psi_2(\mathbf{z}))=d\phi_s(\psi_2(\mathbf{z}))/d\psi_2(\mathbf{z})$, for $s\in \{1, \ldots, m\}$,
which turns out to be instrumental to write the HiMod counterpart of both the diffusive and advective contributions in \eqref{a_ADR}.
Let us deal with these two terms, separately. Concerning the diffusion part, we can expand it as
\begin{equation}\label{HiMod_D}
\begin{array}{lll}
&&\displaystyle\int_{\Omega} \nu(\mathbf{z})\ \nabla u_m(\mathbf{z}) \cdot \nabla v_m(\mathbf{z}) d \mathbf{z}\\[3mm]
&=& \displaystyle
\sum_{j=1}^{m} \sum_{i=1}^{N_h} u_{ij} \displaystyle \int_{\Omega} \nu(\mathbf{z})\Big\{ \big[ \big([\partial_x \psi_1(\mathbf{z})]^2 + |\nabla_{{\mathbf{y}}} \psi_1(\mathbf{z})|^2 \big)\phi_j \phi_k\big]\mathcal{R}'_i \mathcal{R}'_l\\[5mm]
&+& \big[ \big( \partial_x \psi_1(\mathbf{z}) \partial_x \psi_2(\mathbf{z}) + \nabla_{{\mathbf{y}}} \psi_1(\mathbf{z}) \cdot \nabla_{{\mathbf{y}}} \psi_2(\mathbf{z})\big)\phi_j \phi'_k \big] \mathcal{R}'_i \mathcal{R}_l\\[3mm]
&+& \big[ \big( \partial_x \psi_1(\mathbf{z}) \partial_x \psi_2(\mathbf{z}) + \nabla_{{\mathbf{y}}} \psi_1(\mathbf{z}) \cdot \nabla_{{\mathbf{y}}} \psi_2(\mathbf{z})\big)\phi'_j \phi_k \big] \mathcal{R}_i \mathcal{R}'_l\\[3mm]
&+&\big[ \big([\partial_x \psi_2(\mathbf{z})]^2 + |\nabla_{{\mathbf{y}}} \psi_2(\mathbf{z})|^2 \big)\phi'_j \phi'_k\big]\mathcal{R}_i \mathcal{R}_l
\Big\}\, d \mathbf{z},
\end{array}
\end{equation}
where the dependence of $\mathcal{R}_i$
and $\mathcal{R}_l$ on $\psi_1(\mathbf{z})$ as well as of $\phi_j$ and $\phi_k$ on $\psi_2(\mathbf{z})$ is understood.
In a similar way, the convective term in \eqref{a_ADR} becomes
\begin{equation}\label{HiMod_A}
\begin{array}{lll}
&&\displaystyle\int_{\Omega} \mathbf{b}(\mathbf{z}) \cdot \nabla u_m(\mathbf{z}) v_m(\mathbf{z}) d \mathbf{z} \\[3mm]
&=& \displaystyle
\sum_{j=1}^{m} \sum_{i=1}^{N_h} u_{ij} \displaystyle \int_{\Omega} \Big\{
\big[ \big( b_1(\mathbf{z}) \partial_x \psi_1(\mathbf{z}) + \textbf{b}_2(\mathbf{z})\cdot \nabla_{{\mathbf{y}}} \psi_1(\mathbf{z}) \big) \phi_j \phi_k \big] \mathcal{R}'_i \mathcal{R}_l\\[5mm]
&+& \big[ \big( b_1(\mathbf{z}) \partial_x \psi_2(\mathbf{z}) + \textbf{b}_2(\mathbf{z})\cdot \nabla_{{\mathbf{y}}} \psi_2(\mathbf{z}) \big) \phi'_j \phi_k \big] \mathcal{R}_i \mathcal{R}_l
\Big\}\, d \mathbf{z},
\end{array}
\end{equation}
where the advective field $\mathbf{b}$
is decomposed according to the separation of variables supporting the HiMod reduction, i.e., as $\mathbf{b}(\mathbf{z})=(b_1(\mathbf{z}), \textbf{b}_2(\mathbf{z}))^T$.
Now, by collecting the corresponding terms in \eqref{HiMod_D} and \eqref{HiMod_A} and by exploiting the separation of variables underlying a HiMod reduction together with the map linking the physical with the reference domain, we can rewrite the bilinear form $a(u_m, v_m)$ in \eqref{reduced problem} in a compact way as
\begin{equation}\label{stiff}
\sum_{j=1}^{m}\sum_{i=1}^{N_h} \mathcal{A}_{jk}(\mathcal{R}_i,\mathcal{R}_l){u}_{ij},
\end{equation}
for $k\in \{1, \ldots, m\}$, $l\in \{1, \ldots, N_h\}$, where
\begin{equation}
\begin{array}{rcl}
    \mathcal{A}_{jk}(\mathcal{R}_i,\mathcal{R}_l) & = & \displaystyle \int_{\hat{\Omega}_{1D}} \Big[ \hat{\mathcal{Q}}_{jk}^{11}(\hat x) \,
    \mathcal{R}'_i (\hat x) \mathcal{R}'_l (\hat x) + \hat{\mathcal{Q}}_{jk}^{10} (\hat x) \,
    \mathcal{R}'_i (\hat x) \mathcal{R}_l (\hat x)\\[4mm]
    &+&
    \hat{\mathcal{Q}}_{jk}^{01} (\hat x) \,
    \mathcal{R}_i (\hat x) \mathcal{R}'_l (\hat x) +
    \hat{\mathcal{Q}}_{jk}^{00} (\hat x) \,
    \mathcal{R}_i (\hat x) \mathcal{R}_l (\hat x) \Big]\, d {\hat x},
    \end{array}
\end{equation}
with
\begin{equation}
    \hat{\mathcal{Q}}_{jk}^{st}(\hat{x}) = \displaystyle \int_{\hat{\gamma}} \mathcal{Q}^{st}_{jk}(\hat{\mathbf{z}}) | \text{det}(\mathcal{J}_{{\Phi}})| \, d \hat{\textbf{y}}  \qquad s,t = 0,1,
\end{equation}
\begin{equation}
\begin{array}{lll}
\mathcal{Q}^{11}_{jk}(\hat{\mathbf{z}})&=&
\nu(\Psi^{-1}({\hat{\mathbf{z}}})) \big([\mathcal{D}_1^x(\hat{\mathbf{z}})]^2 + |\mathcal{D}_1^{\mathbf{y}}(\hat{\mathbf{z}})|^2 \big)\phi_j(\hat{\textbf{y}}) \phi_k(\hat{\textbf{y}});\\[3mm]
\mathcal{Q}^{10}_{jk}(\hat{\mathbf{z}})&=& \nu(\Psi^{-1}({\hat{\mathbf{z}}})) \big( \mathcal{D}_1^x(\hat{\mathbf{z}})\, \mathcal{D}_2^x(\hat{\mathbf{z}})  +
\mathcal{D}_1^{\mathbf y}(\hat{\mathbf{z}})\cdot
\mathcal{D}_2^{\mathbf y}(\hat{\mathbf{z}})\big) \phi_j(\hat{\textbf{y}}) \phi'_k(\hat{\textbf{y}})
 \\[3mm]
 &+& \big(b_1(\Psi^{-1}({\hat{\mathbf{z}}}))\mathcal{D}_1^x(\hat{\mathbf{z}}) +
{\mathbf b}_2(\Psi^{-1}({\hat{\mathbf{z}}})) \cdot \mathcal{D}_1^{\mathbf y}(\hat{\mathbf{z}})\big)\phi_j(\hat{\textbf{y}}) \phi_k(\hat{\textbf{y}});\\[3mm]
\mathcal{Q}^{01}_{jk}(\hat{\mathbf{z}})&=& \nu(\Psi^{-1}({\hat{\mathbf{z}}})) \big( \mathcal{D}_1^x(\hat{\mathbf{z}})\, \mathcal{D}_2^x(\hat{\mathbf{z}})  +
\mathcal{D}_1^{\mathbf y}(\hat{\mathbf{z}})\cdot
\mathcal{D}_2^{\mathbf y}(\hat{\mathbf{z}})\big) \phi'_j(\hat{\textbf{y}}) \phi_k(\hat{\textbf{y}});
 \\[3mm]
\mathcal{Q}^{00}_{jk}(\hat{\mathbf{z}})&=& \nu(\Psi^{-1}({\hat{\mathbf{z}}})) \big( [\mathcal{D}_2^x(\hat{\mathbf{z}})]^2 + |\mathcal{D}_2^{\mathbf{y}}(\hat{\mathbf{z}})|^2
\big) \phi'_j(\hat{\textbf{y}}) \phi'_k(\hat{\textbf{y}})\\[3mm]
 &+& \big(b_1(\Psi^{-1}({\hat{\mathbf{z}}}))\mathcal{D}_2^x(\hat{\mathbf{z}}) +
{\mathbf b}_2(\Psi^{-1}({\hat{\mathbf{z}}})) \cdot \mathcal{D}_2^{\mathbf y}(\hat{\mathbf{z}})\big)\phi'_j(\hat{\textbf{y}}) \phi_k(\hat{\textbf{y}}),
\end{array}
\end{equation}
and where we have introduced the
deformation indices
$$
\begin{array}{c}
\mathcal{D}_1^x(\hat{\mathbf{z}})=\partial_x \psi_1({\mathbf{z}})\big|_{{\mathbf{z}}=\Psi^{-1}(\hat{\mathbf{z}})},\qquad
\mathcal{D}_1^{\bf y}(\hat{\mathbf{z}})=\nabla_{\bf y} \psi_1({\mathbf{z}})\big|_{{\mathbf{z}}=\Psi^{-1}(\hat{\mathbf{z}})},\\[3mm]
\mathcal{D}_2^x(\hat{\mathbf{z}})=\partial_x \psi_2({\mathbf{z}})\big|_{{\mathbf{z}}=\Psi^{-1}(\hat{\mathbf{z}})},\qquad\mathcal{D}_2^{\bf y}(\hat{\mathbf{z}})=\nabla_{\bf y} \psi_{2}({\mathbf{z}})\big|_{{\mathbf{z}}=\Psi^{-1}(\hat{\mathbf{z}})}
\end{array}
$$
coinciding with the components of the Jacobian $\mathcal{J}_\Psi({\mathbf{z}})$ evaluated at ${\mathbf{z}}=\Psi^{-1}(\hat{\mathbf{z}})$. Coefficients $\hat{\mathcal{Q}}_{jk}^{st}$ synthesize the dynamics transverse to the centerline. They explicitly depend on the problem data, on the geometry of the domain $\Omega$ (through the deformation indices) and on the selected modal basis.

Concerning the linear form $F(v_m)$ on the right-hand side in \eqref{reduced problem}, we apply the same steps used to manipulate the bilinear form $a(u_m, v_m)$, to obtain
\begin{equation}\label{rhs}
\displaystyle \int_{\hat{\Omega}_{1D}} {\hat{\mathcal M}}_k(\hat x)\, \mathcal{R}_l (\hat x)\, d{\hat x}
\end{equation}
for $k\in \{1,\ldots,m\}$, $l\in \{1,\ldots,N_h\}$, with
$$
{\hat{\mathcal M}}_k(\hat x)= \displaystyle \int_{\hat{\gamma}} \mathcal{M}_k(\hat{\mathbf{z}}) | \text{det}(\mathcal{J}_{{\Phi}})| \, d \hat{\textbf{y}}\quad \mbox{and}\quad
\mathcal{M}_k(\hat{\mathbf{z}})=f(\Psi^{-1}({\hat{\mathbf{z}}})) \phi_k(\hat{\textbf{y}}).
$$

Moving from \eqref{stiff} and \eqref{rhs} and by varying indices $k$ and $l$ in $\{1, \ldots, m\}$ and $\{1, \ldots, N_h\}$, respectively we commute the HiMod formulation \eqref{reduced problem} into a system of linear equations of order $mN_h$, with unknowns the $mN_h$ coefficients $\{ u_{ij}\}$. We refer to~\cite{perotto2010hierarchical} for more details about the blockwise structure of the stiffness matrix associated with \eqref{stiff}.

\begin{remark}
\label{rem:lifting}
The term associated with the lifting in \eqref{F_ADR} modifies the HiMod right-hand side in \eqref{rhs} with an additional contribution. In particular, the better way to proceed leads us to expand the lifting $R_{u_{in}}$ in terms of the NURBS functions and of the modal basis according to \eqref{expanded}, so that
$$
R_{u_{in}}(\mathbf{z}) = \sum_{j=1}^{m} \sum_{i=1}^{N_h} r_{ij} \mathcal{R}_i(\psi_1(\mathbf{z}))\phi_j(\psi_2(\mathbf{z})).
$$
For this choice, the HiMod counterpart of the bilinear form $-a(R_{u_{in}}, v)$ in \eqref{F_ADR} simplifies to
$$
\sum_{j=1}^{m}\sum_{i=1}^{N_h} \mathcal{A}_{jk}(\mathcal{R}_i,\mathcal{R}_l){r}_{ij},
$$
with $k\in \{1,\ldots,m\}$, $l\in \{1,\ldots,N_h\}$.
\end{remark}

\section{Numerical results and discussion}
\label{sec:res}
In this section we apply the HiMod discretization to model the transport of a passive scalar in an S-shaped microchannel, by hierarchically reducing
problem \eqref{problem_adr3d} in the computational domain $\Omega$ displayed in Figure \ref{fig1:Methods}.

The geometry of the microchannel is characterized by a length L of $1410$ $\mu$m, with a circular section of diameter $D = 100$ $\mu$m, a first radius of curvature $R_1$ equal to $455$ $\mu$m and a second radius of curvature $R_2$ equal to $204$ $\mu$m.

Concerning the problem data,
we assume that the external force $f$ in \eqref{problem_adr3d} is null, while the convective field $\mathbf{b}$ is obtained by solving the steady-state Navier-Stokes equations in $\Omega$. To this aim, we use the general purpose CFD code Fluent (ANSYS Inc., Canonsburg, PA, \texttt{www.ansys.com}), based on a finite volume discretization. The Navier-Stokes simulation is completed with the following boundary data: we prescribe a constant velocity of $2$ $m/s$ at the inlet section $\Gamma_{in}$; a traction-free condition at the outlet section $\Gamma_{out}$; a no-slip condition on $\Gamma_{w}$, thus assuming the microchannel wall to be rigid.
The dynamic viscosity of the fluid is set to $\mu = 10^{-3}$ $kg/ms$, while the density is $\rho = 998$ $kg/m^3$.
Therefore, the resulting regime for the fluid flow is laminar, as the Reynolds number at the inlet section is approximately equal to $200$.

The diffusion coefficient of the transported species in  \eqref{problem_adr3d} is $\nu=10^{-6}$ $m^2/s$, so that the P\'eclet number at the inlet section corresponds to ${\mathbb P}{\mbox e} = 200$.
Finally, as for the boundary conditions of the AD equation, the scalar distribution $u_{in}$ corresponds to a parabolic profile on the inlet section $\Gamma_{in}$, while we assign a homogeneous Dirichlet boundary condition on the microchannel wall $\Gamma_w$, and a homogeneous Neumann data on the outlet section $\Gamma_{out}$.
\begin{figure}
   \centering
    \includegraphics[width=12cm]{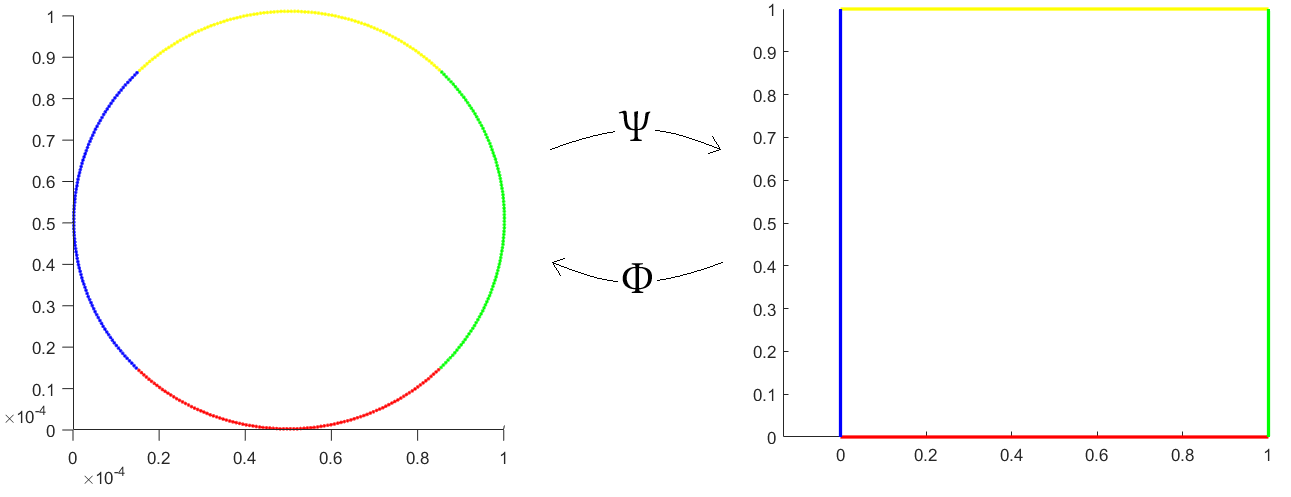}
    \caption{Correspondence
    between boundary portions of
    the generic transverse fiber and of the reference unit square.}
    \label{phi_psi2D}
\end{figure}

\begin{figure}
   \centering
    \includegraphics[width=12cm]{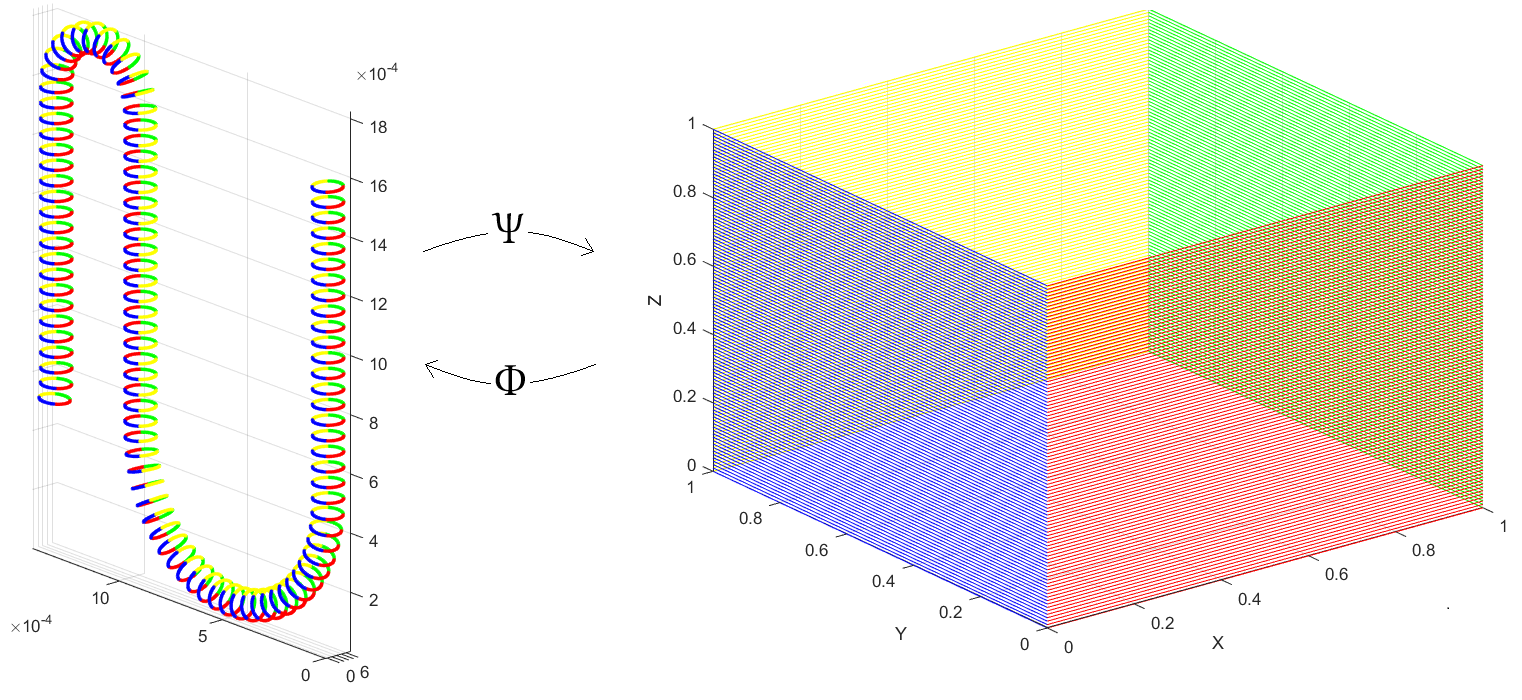}
    \caption{Sketch of the maps $\Phi$ and $\Psi$ between the S-shaped microchannel and the reference unite cube.}
    \label{phi_psi3D}
\end{figure}

Concerning the geometric characterization of the computational domain in view of a HiMod reduction, the leading direction is aligned to the centerline of the microchannel (i.e., to the direction the convective field flows along), while the transverse sections coincide with the uniform circular cross-sections of the channel. \\
The NURBS description of volume $\Omega$ is built as follows. For each point $\mathbf{P}=(x_p, \textbf{y}_p)$ along the centerline (i.e., such that $x_p\in \Omega_{1D}$), we provide the function $\mathbf{p}: [0, 2\pi] \to \mathbb{R}^3$ such that the curve
\begin{equation}
    \mathbf{p}(\theta)=\frac{D}{2} \sin(\theta)\mathbf{B}+\frac{D}{2}\cos(\theta)\mathbf{N}+\mathbf{P}
    \label{wall_section}
\end{equation}
identifies the boundary of the transverse fiber, $\gamma_{x_P}$, of $\Omega$ centered at $\mathbf{P}$, where
vectors $\mathbf{B}$ and $\mathbf{N}$ correspond to the binormal and to the normal unit vectors of the Frenet frame at $\mathbf{P}$, respectively. Such vectors can be easily computed by using, e.g., \texttt{vmtk} (\texttt{www.vmtk.org}).\\
Now, we focus on maps $\Phi$ and $\Psi$, relating the physical with the reference domain. First, we introduce
a NURBS map between each side of the reference transverse fiber $\hat{\gamma} = (0, 1)^2$ and a portion of the wall of the generic circular cross section.
To this aim, the curve defined by \eqref{wall_section} is built by varying angle $\theta$ in the four disjoint intervals $[-\pi/4, \pi/4)$, $[\pi/4, 3/4 \pi)$, $[3/4 \pi, 5/4 \pi)$, $[5/4 \pi, 7/4 \pi)$. Successively, each of the four resulting curves is mapped to one side of the unit square $\hat{\gamma}$, according to the correspondence highlighted in Figure \ref{phi_psi2D}.
Afterwards, a NURBS parameterization of the internal part of the transverse fiber is carried out by means of a bilinearly blended Coons patch (i.e., via a suitable bilinear interpolation of the four arcs in Figure~\ref{phi_psi2D}, see, e.g.,~\cite{farin1999discrete}), as implemented in the Matlab NURBS toolbox.
Finally, maps $\Phi$ and $\Psi$ are obtained by repeating the procedure above for each point $\mathbf{P}$ along the centerline, as sketched in Figure~\ref{phi_psi3D}.

\begin{figure}
   \centering
    \includegraphics[width=12cm]{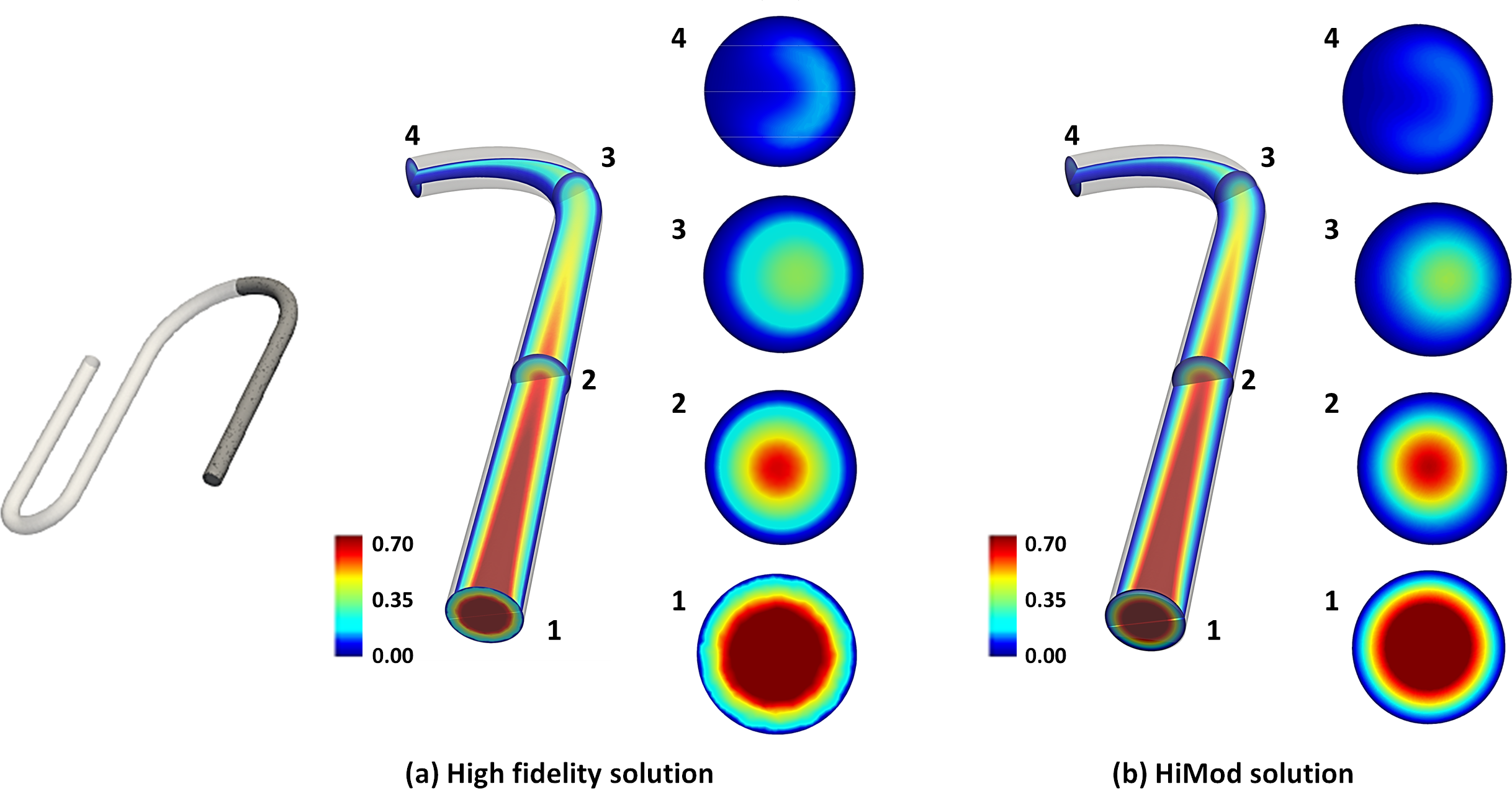}
    \caption{Qualitative comparison between the high fidelity solution and the HiMod approximation in correspondence of the microchannel's portion which is highlighted on the left: cut-view of the colourplot of the passive scalar concentration in the plane containing the centerline and at four different sections of interest.}
    \label{confronti_interi}
\end{figure}

To identify the discrete HiMod space in \eqref{global function space}, we resort to a modal basis consisting of
$64$ functions which describe the dynamics parallel to the transverse direction, while we discretize the flow aligned with the centerline of the microchannel by means of quadratic NURBS basis functions, characterized by a $C^1$ inter-element smoothness, after introducing $202$ uniformly-spaced knots along the reference supporting fiber $\hat \Omega_{1D}$.
Panel (b) in Figure~\ref{confronti_interi} shows the HiMod solution. Four cross sections of interest are highlighted: the inlet (\#1), a section halfway through the first straight part of the microchannel (\#2), a section at the beginning of the region with curvature $R_1$ (\#3), and finally a section halfway through the region with curvature $R_1$ (\#4).
Furthermore, a cut in the plane which contains the centerline is also shown.

For comparison purposes, we run a simulation of the same AD problem with Fluent on a uniform unstructured computational grid of $\Omega$ consisting of $1.384.925$ tetrahedral elements and $273.421$ vertices. In particular, Fluent solves the  Navier-Stokes and the AD equations concurrently. We adopt a second-order discretization scheme for the pressure, and a second-order upwind method to approximate both the velocity and the passive scalar concentration.
The resulting solution is considered as the high fidelity reference solution, and is employed to establish the accuracy of the HiMod approximation.
\begin{figure}
   \centering
    \includegraphics[width=8cm]{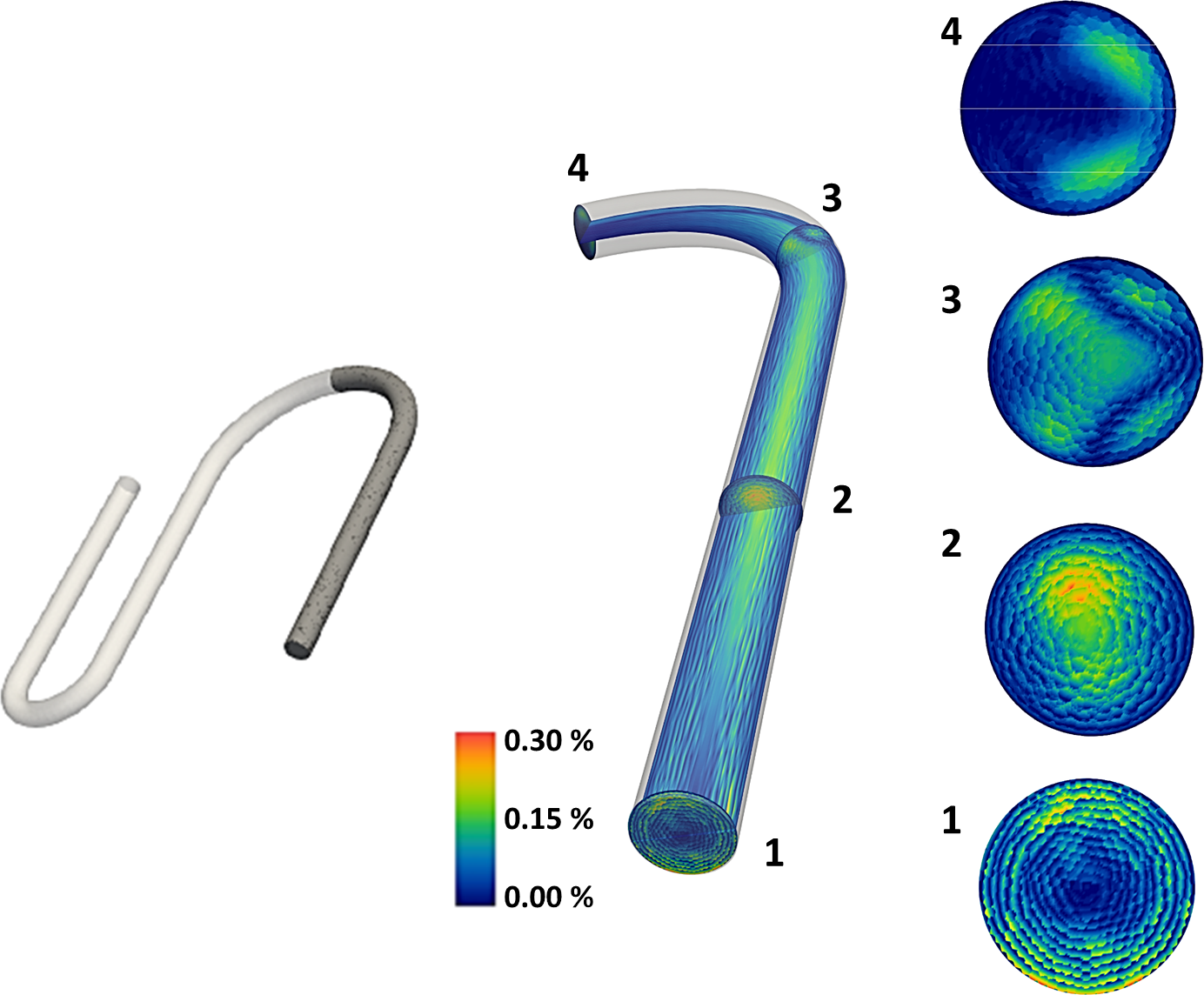}
    \caption{Quantitative comparison between the high fidelity solution and the HiMod approximation in correspondence of the microchannel's portion which is highlighted on the left: pointwise distribution of the $L^2(\Omega)$-norm of the relative modeling error between the two models.}
    \label{errore_assoluto}
\end{figure}

A qualitative comparison between the HiMod and the reference solution is displayed in Figure~\ref{confronti_interi}. We observe
a good qualitative agreement between the two solutions. We highlight that the data assigned at the inlet of the HiMod reduced model does not exactly coincide with the parabolic profile $u_{in}$ used in the reference context.
Indeed, as discussed in Remark~\ref{rem:lifting}, the lifting $R_{u_{in}}$ in \eqref{F_ADR} is built upon the expansion of the Dirichlet data $u_{in}$ in terms of the modal and of the NURBS bases.
This justifies the minor discrepancy
in the concentration on the inlet section (\#1). Such a slight mismatch is unavoidably propagated along the channel and explains the small error
present on the downstream sections.
In particular, the HiMod solution slightly overestimates the high fidelity concentration on section \#2, while underestimation occurs on sections \#3 and \#4.

A more quantitative comparison is performed in Figure~\ref{errore_assoluto}, which displays the pointwise distribution of the relative modeling error between the high fidelity and the reduced solution with respect to the $L^2(\Omega)$-norm, in correspondence with the plane and the four sections considered in Figure~\ref{confronti_interi}.
We can infer that a small relative error (less than $1\%$) is achieved, which is often acceptable for the type of engineering applications analyzed in this chapter.\\
Finally, from a computational viewpoint, the gain provided by the HiMod approximation is confirmed by the number of the degrees of freedom (dofs) characterizing the two solutions, namely $1.384.925$ dofs for the high fidelity model to be compared with $12.928$ dofs for the reduced model.

\section{Conclusions and perspectives}
\label{sec:concl}
An S-shaped microchannel used to enhance mixing by yielding chaotic advection in passive micromixers provides the ideal environment where to verify the computational performances of a HiMod reduction procedure, both in terms of reliability and computational efficiency.
Indeed, it is common to distinguish in the microchannel geometry a leading direction, aligned with the channel centerline, and an orthogonal transverse direction, parallel to the cross section, coherently with the separation of variables exploited in the definition of the HiMod reduced space.\\
The intrinsic capability of Himod reduction to decouple the dominant dynamic from the secondary one through a differentiated discretization of the two directions leads to the resolution of a system characterized by a considerably lower number of unknowns, without quitting the accuracy of the reduced solution.
In particular, for a $99.07\%$ reduction in the number of dofs with respect to a high fidelity finite volume simulation, a very good accuracy is exhibited by the HiMod solution, with a less than $1\%$ relative modeling error.

In a future perspective, the proposed approach may be extended to more complex problems in the microfluidics field, such as the study of AD problems under unsteady conditions or in the presence of more complex 3D geometries. Moreover, a model reduction approach could speed up the microchannel design phase, especially when different scenarios have to be investigated in order to optimize the mixing efficiency of the microsystem. The parametric version of the HiMod reduction could be instrumental for such a goal (\cite{Barolietal17,Zanca21,lupopasini2022}).

\section*{Acknowledgments}
All the authors acknowledge the European Union’s Horizon 2020 research and innovation programme under the Marie Skłodowska-Curie Actions, grant agreement 872442
(ARIA, Accurate Roms for Industrial Applications). Moreover, SP thanks the PRIN research grant n.20204LN5N5
(Advanced Polyhedral Discretisations of Heterogeneous PDEs for Multiphysics Problems);
UM thanks the MIUR
FISR   FISR2019\_ 03221 CECOMES project;
FB thanks the project ``Reduced order modelling for numerical simulation of partial differential equations'' funded by Università Cattolica del Sacro Cuore.

\bibliography{references.bib}
\end{document}